\documentclass[manuscript]{aastex6}

\usepackage[varg]{txfonts}
\usepackage{natbib}
\usepackage{color}
\usepackage{graphicx}
\usepackage{hyperref}
\usepackage{epstopdf}
\usepackage{gensymb}
\usepackage{epsfig}
\usepackage{subfigure}
\usepackage{fancyhdr}

\begin{document}
\title{\bf{First Imaging Observation of Standing Slow Wave in Coronal Fan loops}}
\author{V.~Pant$^{1}$,
        A. Tiwari$^{1,2}$,
        D. Yuan$^{3}$
        D. Banerjee$^{1,4}$
        }
\affil{$^{1}$ Indian Institute of Astrophysics, Bangalore-560 034, India\\
         $^{2}$ Northumbria University, Newcastle Upon Tyne, NE1 8ST, UK\\
         $^{3}$ Institute of Space Science and Applied Technology, Harbin Institute of Technology, Shenzhen 518000, China\\
           $^{4}$Center of Excellence in Space Sciences, IISER Kolkata, India}

\begin{abstract}
 We observe intensity oscillations along coronal fan loops associated with the active region AR 11428. The intensity oscillations were triggered by blast waves which were generated due to X-class flares in the distant active region AR 11429. 
 To characterise the nature of oscillations, we created time--distance maps along the fan loops and noted that the intensity oscillations at two ends of the loops were out of phase. As we move along the fan loop,  the amplitude of the oscillations first decreased and then increased. The  out--of--phase nature together with the amplitude variation along the loop implies that these oscillations are very likely to be standing waves. The period of the oscillations are estimated to be $\sim$27 min, damping time to be $\sim$45 min and phase velocity projected in the plane of sky $\sim$ 65--83 km s$^{-1}$. The projected phase speeds were in the range of acoustic speed of coronal plasma at about 0.6 MK which further indicates that these are slow waves. To best of our knowledge, this is the first report on the existence of the standing slow waves in non--flaring fan loops. 
\end{abstract}
\newpage


\keywords{Sun: Corona ; Sun: Coronal loops ; Sun: Oscillations}

\section{Introduction}
Magnetohydrodynamic (MHD) waves are ubiquitous in solar corona. With the advent of modern space based instruments, different types of wave modes have been observed in the last decade. 
Slow MHD modes (compressional waves) were first observed in the polar coronal holes using UVCS by \citet{1997ApJ...491L.111O}. Later, \citet{1998ApJ...501L.217D} and \citet{1999ApJ...514..441O} reported propagating intensity disturbances (PDs) in polar plumes using EIT onboard SOHO. 
Recently, several authors have reported that small scale jets and spicules at transition region and chromosphere are associated with with PDs seen in polar plumes and polar coronal holes \citep{2015ApJ...807...71P,2015ApJ...809L..17J,2015ApJ...815L..16S,2016ApJ...829L..18B,2016ApJS..224...30Y}.  Reflections of propagating slow waves were also reported in hot and flaring coronal loops using AIA \citep{2013ApJ...779L...7K,2015ApJ...804....4K} and XRT observations \citep{2016ApJ...828...72M}. The authors have reported that these waves are triggered by the flares at the footpoint of the coronal loops. Recently, \citet{2015ApJ...813...33F} have modelled reflective slow mode in flaring loops using 2.5D magnetohydrodynamic simulation in synthetic 131~\AA~emission images.

Apart from propagating slow waves, flare--excited standing slow waves have also been observed in hot and flaring coronal loops. Oscillations in Doppler velocity, detected in Fe {\sc xix}, were reported in hot flaring coronal loops using SUMER/SOHO and SXT/Yokoh \citep{2002ApJ...574L.101W}. Time period of oscillations was found to be 14--18 min. These oscillations were interpreted as slow standing modes. \citet{2003A&A...406.1105W,2003A&A...402L..17W} have performed statistical study of slow standing modes in several hot coronal loops and post flare loops, respectively. They have reported a $\pi$/2 phase shift between Doppler velocities and line intensities of Fe  {\sc xix} and Fe {\sc xxi} emission lines (formation T $>$ 6 MK) which is the signature of a standing slow modes \citep[see also, ][]{2011SSRv..158..397W,2015ApJ...807...98Y}. \\
The standing slow modes are believed to be triggered by an impulsive flare which cause asymmetric heating at one footpoint of the coronal loop \citep{2005A&A...435..753W}. However, \citet{2004A&A...422..351T} performed a numerical study of the longitudinal oscillations and reported that the excitation of standing oscillations is independent of the location of the impulsive heating in the loop.  \citet{2005A&A...438..713T,2007ApJ...659L.173T} have performed 1D hydrodynamic simulations of standing slow mode and showed that slow standing waves can be triggered by impulsive foot point heating as well. \citet{2007ApJ...659L.173T,2008A&A...481..247T} have constructed 1D hydrodynamic loop model to study and distinguish between standing and propagating slow oscillations in hot and cool coronal loops. They have reported that the phase of the intensity of the oscillation continuously changes with time due to heating and cooling of loops. \citet{2015ApJ...807...98Y} have performed forward modelling of standing slow modes in hot flaring coronal loops (T $>$ 6 MK) and studied their imaging and spectroscopic signatures. The authors have reported that the amplitude of the oscillations along the loop should vary depending on the mode of the oscillations. \\
Slow standing waves are found to be strongly damped. \citet{2002ApJ...580L..85O} used a 1D MHD model to study damped standing slow oscillations. The strong damping was attributed to large thermal conduction which depends on the temperature of the loops. Recently, \citet{2015ApJ...811L..13W} observed standing slow modes in hot coronal loops using AIA 94~\AA~observations and reported that the thermal conduction, which is believed to damp the standing oscillations, is suppressed in hot coronal loops. Till today, standing slow modes have been observed exclusively in hot coronal loops. A very limited number of standing slow mode waves were detected by imaging observations. In this letter, we report the evidence of standing slow waves in cool fan loops. The paper is organised as follows. In section~\ref{obs}, we describe the data processing used for this study. In section~\ref{data}, we describe the method of analysis which is followed by discussion and conclusions in section~\ref{discuss}
\clearpage
\section{Observations}
\label{obs}
\begin{figure*}[!h]
\centering
\includegraphics[scale=0.75,angle=90]{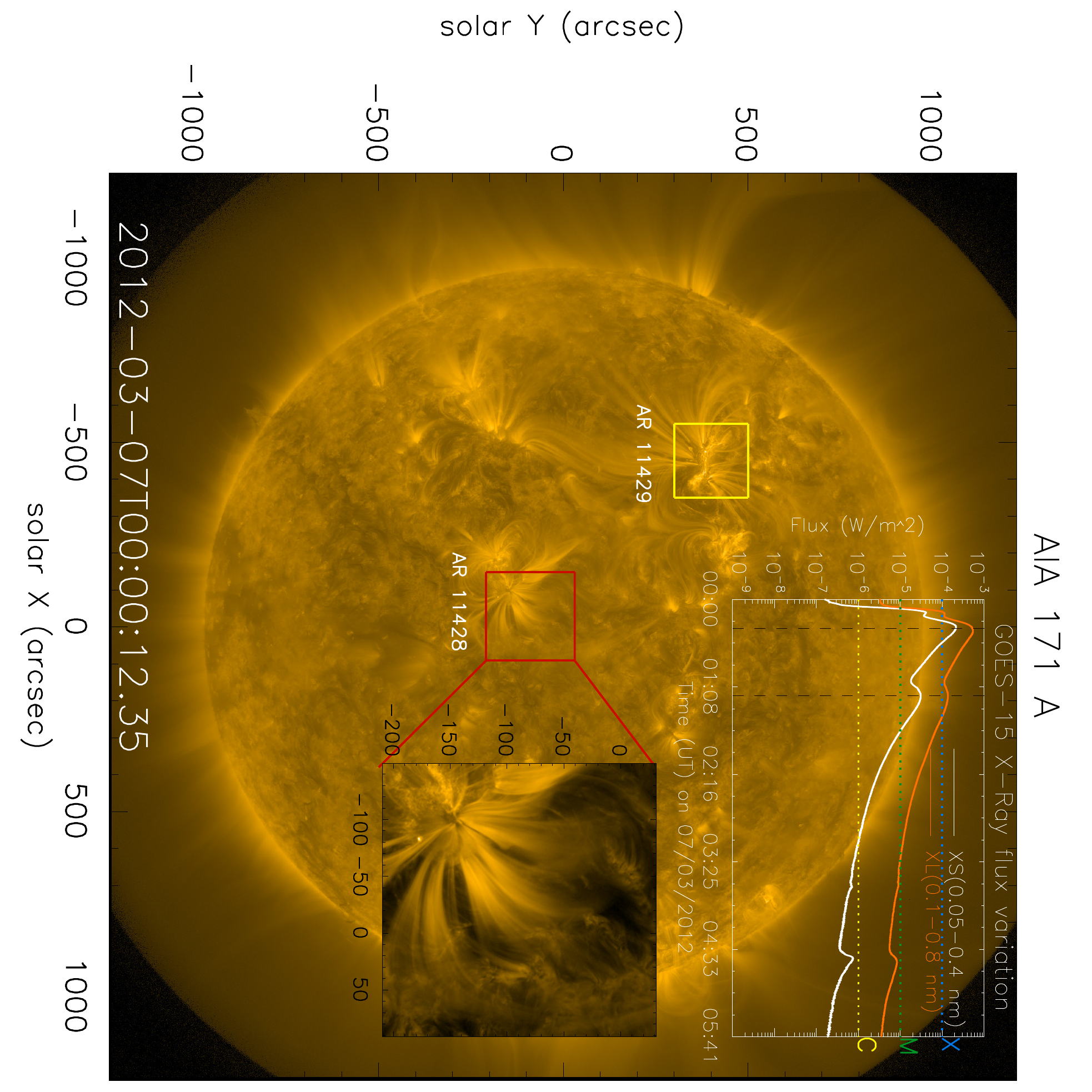}
\caption{Full disk image of Sun at AIA 171~\AA. Red and yellow box represent the location of active regions AR 11428 and 11429 respectively. Region of interest (ROI) which is used for further analysis is enclosed in the box shown in red. GOES X--ray flux variation is overplotted in the figure. Curves in orange and white represents the flux corresponding to two passbands, {\it i.e,} 0.1--0.8 nm and 0.05--0.4 nm, respectively. Two vertical dashed lines in black represent the timings of the peak of the GOES X-ray flux relevant for this study. }
\label{fig1} 
\end{figure*}

On 2012 March 7, a group of fan loops were observed near the active region AR 11428 (see Figure~\ref{fig1}). Two X-class flares were detected consecutively at a distant active region, AR 11429 to the north west of AR 11428. The approximate distance between AR 11428 and AR 11429 is about 455 Mm. The GOES X-ray emission (inset in Figure~\ref{fig1}) exhibit the evolution of the flares. The X-ray flux at two channels peaked at 00:22 UT and 01:13 UT, respectively.  The strength of two peaks corresponds to the fluxes of X5 and X1 classes, respectively. Both the X5 and X1 flares originated from AR11429 and the associated energy pulses reach AR11428 at 00:27UT and 01:15 UT, respectively. The fan loops were initially driven to move transversely, and subsequently the intensity perturbations along the loops became detectable. A three-hour data set (00:00 UT-03:00 UT) taken by the Atmospheric Imaging Assembly (AIA) on board Solar Dynamics Observatory (SDO) \citep{lemen2011} was used for detailed analysis. The fan loops of interests are visible in both 171~\AA~and 193~\AA, so we only use these two channels for study. 
\begin{figure*}[!h]
\centering
\includegraphics[scale=0.75,angle=90]{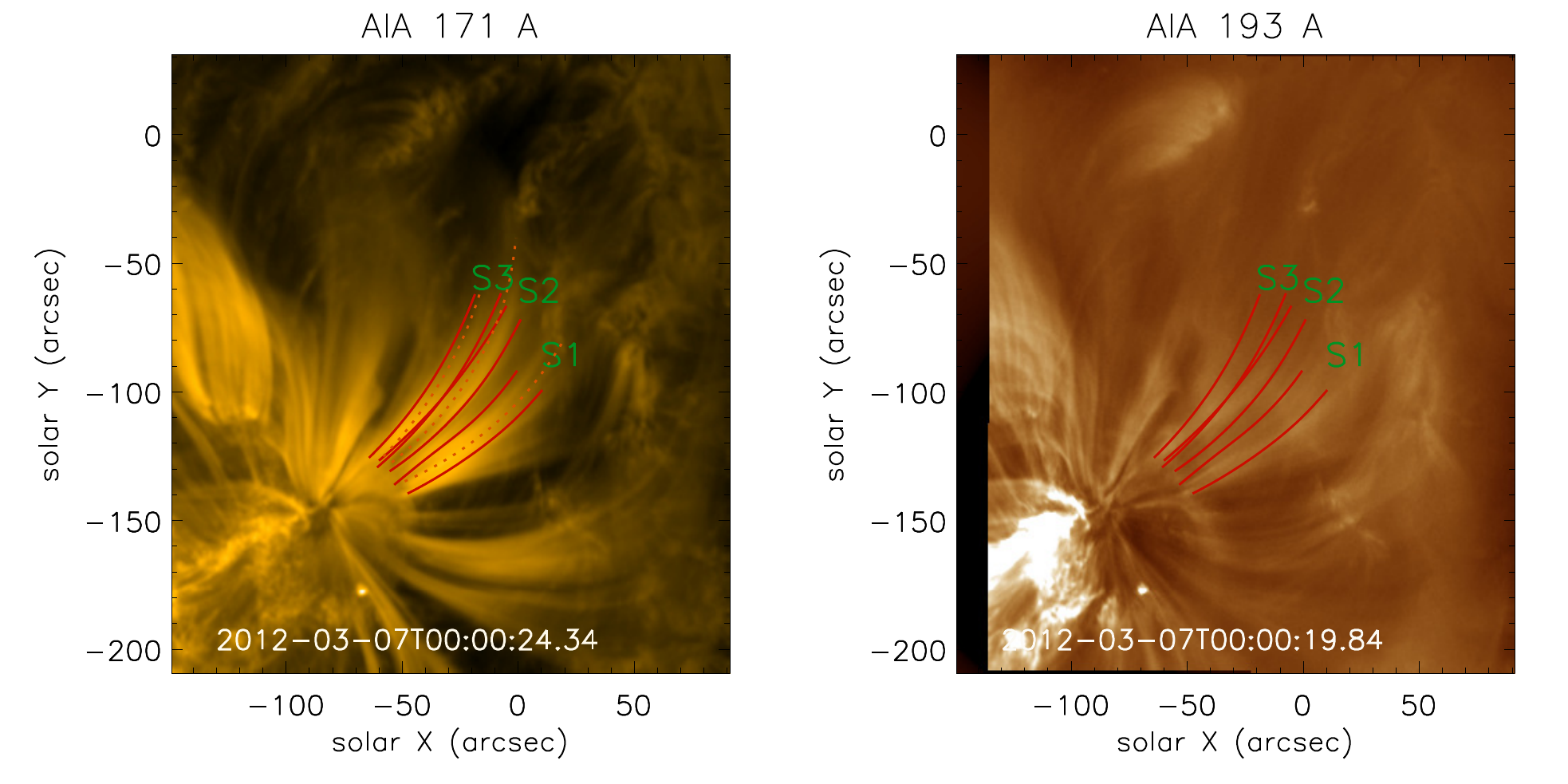}
\caption{ {\it Left}: AIA 171~\AA~ image of the ROI shown in Figure~\ref{fig1}. Three curved artificial broad slices, S1, S2 and S3 are overplotted in red which were used to generate the x-t maps shown in Figure~\ref{fig3}. The dotted curves in orange represents the length of the fan loops. {\it Right}: Same as left but for AIA 193~\AA. Movies~1 and 2 are corresponding to AIA~171 and 193~\AA~ are linked with this figure.}
\label{fig2} 
\end{figure*}

\section{Results}
\label{data}
\subsection{Time evolution of intensity oscillations}
\label{td}
\begin{figure*}[!h]
\centering
\includegraphics[scale=0.95,angle=90]{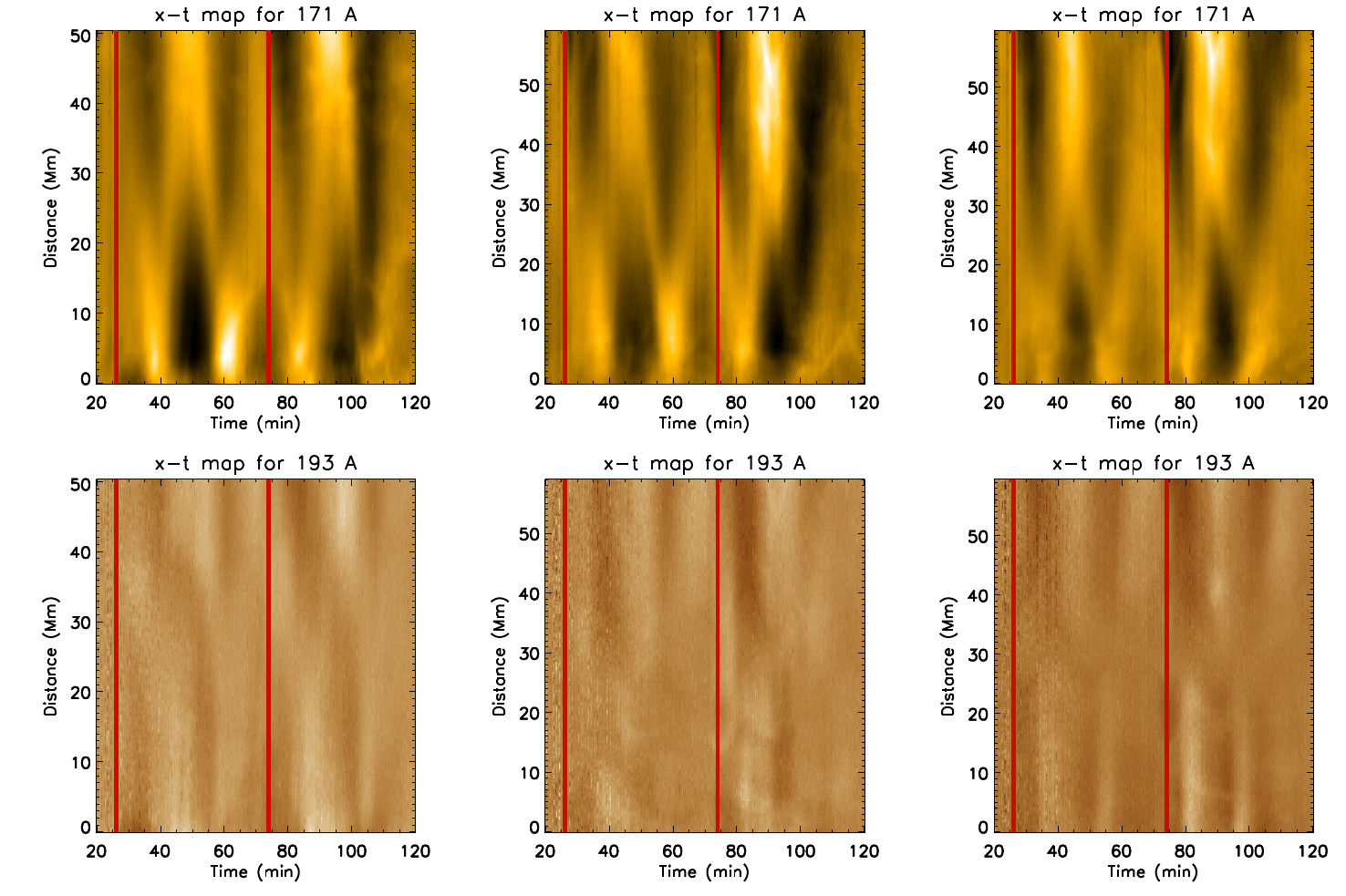}
\caption{Time - distance (x-t) maps corresponding to slices S1, S2 and S3 as marked in Figure~\ref{fig2} are shown in left, middle and right panels, respectively. Two vertical red lines represent the time at which two blast waves impacted the fan loop system. Y-axis represents the distance along the artificial slice.}
\label{fig3} 
\end{figure*}

To derive the properties of the oscillations, we placed three artificial slices, S1, S2, and S3, along the fan loops as shown in Figure~\ref{fig2}, at the locations where the intensity oscillations were clearly seen. We chose broad artificial slices in order to capture the longitudinal oscillations despite they get displaced in transverse direction due to interaction with the blast wave. It should be worth noting at this point that only one footpoint, close to the active region, of fan loops was clearly visible in 171 and 193~\AA. The length of the artificial slices correspond to the distances along the fan loop up to which clear signatures of intensity oscillations were observed. Therefore, the length of the artificial slices may not be equal to the length of the fan loops. We discuss the estimation of the length of the fan loop in section~\ref{loop}. 
For each of the three artificial slices, we generated a time--distance map, which will be termed as an x-t map, henceforth, throughout the paper. Figure~\ref{fig3} represents the x-t maps for slices S1, S2, and
S3 for 171 and 193~\AA~in left, middle and right panels respectively. 
The signatures of intensity oscillations were  clearer in AIA~171~\AA~as compared to AIA~193~\AA, because fan loops appeared more diffuse in AIA~193~\AA. A possible reason for this is discussed in section~\ref{dem}. The vertical lines in red in Figure~\ref{fig3} represent the instances when blast waves hit the fan loops system. We noticed that the second blast wave hit fan loops when the intensity oscillations driven by the first blast wave were still present. 

Figure~\ref{fig3} shows that the intensity oscillations were out of phase at the two ends of the artificial slices as seen in the x-t maps. The out--of--phase signature was clearly seen in both  AIA 171 and 193~\AA. It is clear from the Figure~\ref{fig3} that one reflection point (or antinode) of the oscillations was present near the one footpoint which is clearly visible in the intensity images. While the other antinode was present at the other end of the artificial slice which may or may not be close to the other footpoint. From movies~1 and 2 linked with Figure~\ref{fig2}, it is evident that the shape and appearance of fan loops was changed after the second blast wave hit the fan loop system. Therefore, intensity oscillations were not clearly seen in x-t maps after second blast wave hit the fan loops.\\
Figure~\ref{fig4} shows the variation of intensity with time at different distances along the artificial slice S1. Y-axis represents the relative intensity normalized to the local background. Two dashed vertical lines in red represent the instant of time when blast wave hit fan loops. Since blast wave hit the fan loops twice, we fitted the sinusoidal and damped sinusoidal curve separately at two different time intervals. Red curve represents the best fit sinusoidal curve. We should point out  that the damping of the oscillations were not seen clearly during the first period of observation, which may be due to the impact of the second blast wave. Thus we did not fit damped sinusoidal curve during the first time interval. However, we fitted a damped sinusoidal curve (shown in orange) in the second time interval and noticed the signature of damping at some locations along the fan loops ({\it e.g,} at 5 Mm). The intensity of oscillations became undetectable after 120 min because the shape and appearance of the fan loop changed (see movies~1 and 2). The estimated average period of the oscillation, $P$ and average damping times, $\tau$, at the location of three slices, S1, S2 and S3 in 171 and 193~\AA~are listed in Table~\ref{table1}. Since only one oscillation was observed during second interval, there were large uncertainties in the damping time. The quality factor (ratio of damping time by time period) estimated at the location of three slices is also listed in Table~\ref{table1}. These oscillations are weakly damped as compared to those reported earlier in hot coronal loops. A possible reason for weak damping is outlined in section~\ref{discuss}.

\subsection{Variation of amplitude of intensity oscillation}
\label{amp}
\begin{figure*}[!htb]
\centering
\includegraphics[scale=0.75,angle=90]{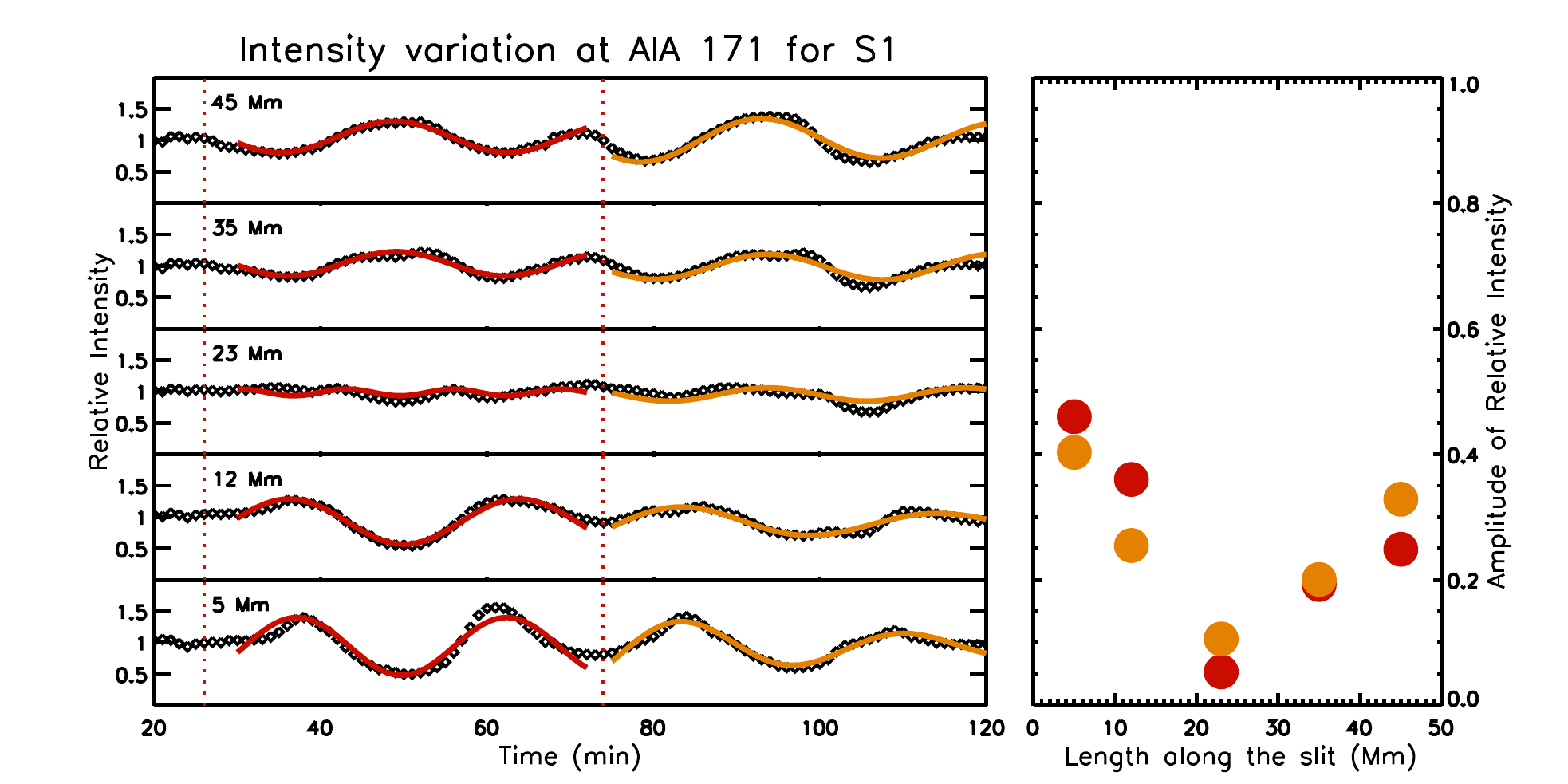}
\includegraphics[scale=0.75,angle=90]{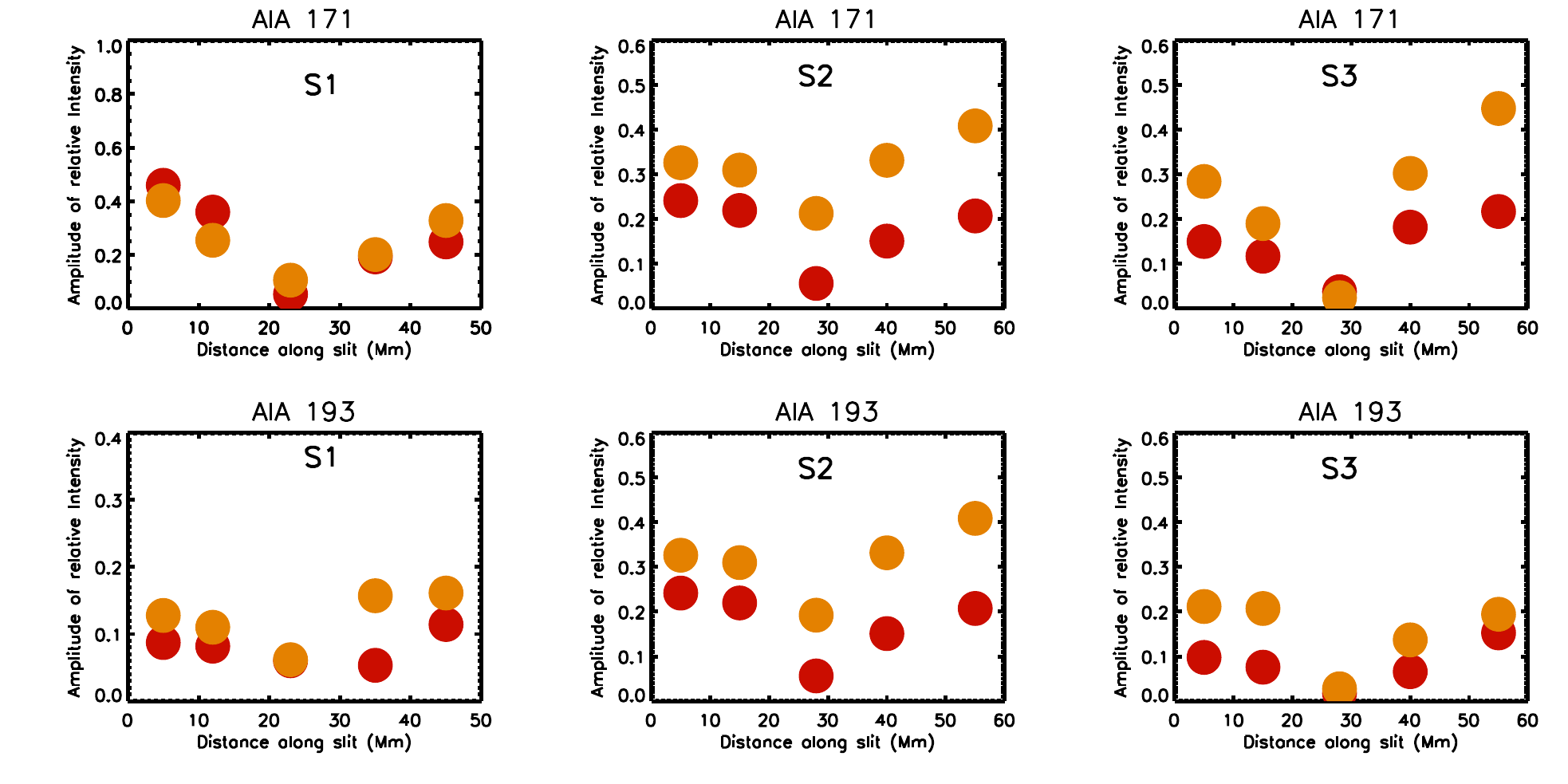}
\caption{{\it Top}: Left:- Intensity variation after normalising to the background intensity, at different distances along S1. Two vertical dashed line represents the instances when blast wave hit the fan loops system. {\it Middle and bottom}: Variation of amplitude of intensity oscillations for S1, S2 and S3 in 171 and 193~\AA. The top row right panel is same as the middle row left panel.}
\label{fig4} 
\end{figure*}

We noted that the relative amplitude (after normalising with background intensity) of the intensity oscillations along S1 in 171~\AA~first decreased and then increased while moving from one end at S1 (close to one footpoint) to other (may be close to another footpoint) (see Figure~\ref{fig4}). The variation can be seen clearly for both curves fitted at two separate time intervals shown in red and orange. Furthermore, the variation of the amplitude at different distances along S1, S2 and S3 in 171 and 193~\AA~is also shown in Figure~\ref{fig4}. Systemic decrease and increase of the amplitude of oscillations, while moving from one end of slice to another, was seen at the location of all slices. This signature clearly indicate the existence of an anti-node near the footpoints of the fan loop. 
\subsection{Estimation of loop length and velocity of the oscillations}
\label{loop}
The footpoint of the fan loop that was away from the active region was distributed and therefore, not seen clearly in normal intensity images of 171 or 193~\AA. Thus, it was not straightforward to measure the length of the fan loops. Moreover, the shape and appearance of fan loops also changed with time (see movie~1 and 2). To estimate the length, we chose the frames where the fan loops were best seen in normal intensity images. We chose several points along the visible segment of the fan loops and interpolated a cubic spline between them. The length of interpolated curve should be approximately equal to the projected length of the fan loops. The orange curves in Figure~\ref{fig2} are the fitted spline curves which represents the projected length of fan loops at the location of three artificial slices. We found the length of the fan loops at the location of S1, S2 and S3 to be  62, 74 and 54 Mm, respectively (see Table~\ref{table1}). Note that the estimated length is the projected length in the plane of sky. Assuming the length of fan loop as same in 171 and 193~\AA, we estimated the phase velocity of oscillations in 171 (193)~\AA~to be 75 (85), 83 (101) and 65 (91)  km~s$^{-1}$ at the location of S1, S2 and S3, respectively. The phase velocity of oscillations, $v$, are comparable to the speed of sound in 171 and 193~\AA.

\subsection{Temperature and density of the fan loop}
\label{dem}
We estimated the temperature and density of the fan loop using automated differential emission measure (DEM) technique as developed by \citet{aschdem}. The temperature of the fan loops was found to be $\sim$ 0.7 MK which are much cooler than the hot loops as observed by hot SUMER lines \citep{2002ApJ...574L.101W} and in the AIA 94 ~\AA~channel \citep{2015ApJ...811L..13W}. We also observed that the electron density decreased along the loop. Since the temperature of fan loops is low, they appear brighter in 171~\AA~channel and diffuse in hotter channels like 193~\AA. We should point out that a fan loop may consists of several finer strands and  we have not considered that scenario here. 

\begin{table}
\begin{center}
\caption{Observational parameters of oscillations}  

\label{table1}
\begin{tabular}{c|ccccc|ccccc}
\hline \hline
\multicolumn{1}{c}{}&\multicolumn{5}{c}{AIA~171~\AA} & \multicolumn{5}{c}{AIA~193~\AA}\\
\hline
Slice & $P$ (min) & $\tau$ (min) & $Q$ & $l$ (Mm) & $v$ (km~s$^{-1}$)& $P$ (min) & $\tau$ (min) & $Q$ & $l$ (Mm) & $v$ (km~s$^{-1}$) \\
S1&27.5$\pm$1.8&40$\pm$25&1.45&62&75&24.1$\pm$5.4&37$\pm$18&0.81&62&85\\
S2&29.6$\pm$3.8 & 53$\pm$25 & 1.79 & 74 & 83&24.4$\pm$4.6&20$\pm$10&1.53&74&101\\
S3&27.6$\pm$4.7& 42$\pm$20  & 1.52 & 54 & 65&19.7$\pm$1.7&42$\pm$12&1.72&54&91\\
\hline

\hline
\end{tabular}
\end{center}
\tablecomments{$P$ represents the period of oscillations, $\tau$ represents the damping time, $Q$ is the quality factor, defined as the ratio of damping time and period of oscillations, $l$ is the projected length of the fan loop at the location of the slice and $v$ is the velocity of the oscillations.}
\end{table}

\section{Discussion and Conclusions}
\label{discuss}
We observe intensity oscillations in a non--flaring fan loop system as seen in AIA~171 and 193~\AA~images. The intensity variations were out of phase close to two footpoints of fan loops and the amplitude of the intensity oscillations varied along fan loops at the location of artificial slices. The amplitude of intensity oscillations first decreased and then increased while moving from one footpoint to another along the fan loop. It should be noted that it is  difficult to identify the differences between standing and propagating waves without spectroscopic signatures. Recently, \citet{2015ApJ...807...98Y} have performed forward modelling of standing slow magnetoacoustic waves in flaring loops. They have reported that the variation of amplitude along the coronal loops is one of the signatures of the standing slow magnetoacoustic  waves \citep[see, Figure~8 in ][]{2015ApJ...807...98Y}. Moreover, small phase shift in the intensity variations with time at different distances along fan loop corresponding to the slice S1, as seen  in Figure~\ref{fig4},  can be due to the presence of standing slow oscillations  \citep{2007ApJ...659L.173T, 2008A&A...481..247T}. We estimated the time period of the oscillations $\sim$27 min and damping time $\sim$45 min. We calculated the projected length of the fan loops and estimated that the velocity of oscillations are comparable to the velocity of sound in 171 and 193~\AA. These signatures allows us to conclude that the observed oscillations are due to standing slow waves in coronal fan loops. The fan loops under study are associated with a sunspot. \citet{2011A&A...533A.116Y}  reported presence of long period oscillations in the coronal diffused plasma near an active region. The oscillations observed in this study are different from those reported by \citet{2011A&A...533A.116Y} because the event under study was triggered by energy impulse of flares, while \citet{2011A&A...533A.116Y} studied persistent leakage of long period oscillations from the underneath sunspot.

It is worth mentioning that only one footpoint of fan loops was clearly seen in AIA 171 and 193~\AA~images. At this stage we can only conjecture two possible scenarios by which the reflection of wave from other end can happen. Either the antinode of the oscillations is present at the other footpoint which is distributed and therefore not seen clearly in normal intensity images or the antinode could be present at the region of sharp density contrast close to the other end of the fan loop. The region of sharp density change may have acted as a reflecting surface. These scenario may be experimented in future studies using computer simulations.

At most of the locations along the fan loops, oscillations are found to be undamped. The reason for the absence of damping at most of the locations is not clear to us, more observations of such events are required to reach conclusive views on the damping. However at few locations along the fan loop, we indeed noted weak damping. The oscillations at those locations are weakly damped as compared to those reported in \citet{2002ApJ...580L..85O,2002ApJ...574L.101W,2003A&A...402L..17W,2015ApJ...811L..13W} where the damping time was comparable to the time period of the oscillations in hot and flaring coronal loops (T $>$ 6 MK). One of the reason for weak damping  could be because the fan loops under study are not hot ($\sim$ 0.7 MK), thus the  thermal conduction may not be efficient enough. Since thermal conduction is one of the main mechanism to damp slow waves, the oscillations were weakly damped in our study.

In summary, we found the signatures of standing slow magnetoacoustic waves in cool fan loops. In earlier studies these oscillations were particularly observed in the hot coronal loops. To the best of our knowledge, this is the first report, on the observational signatures of the existence of weakly damped standing oscillations in cool fan loops.
\section{Acknowledgments}
The authors thank the referee for her/his valuable in-depth comments which have helped us to improve the manuscript.



\end{document}